# Chaos in Mean Motion Resonances of the Kuiper Belt

Fred A Franklin and Paul R Soper
Harvard-Smithsonian Center for Astrophysics

A recent paper of ours [2012, arXiv: 1207.4762] claimed that some level of population density in the outer Kuiper belt, i.e., the sparsely populated region beyond the 1/2 mean motion resonance (mmr) with Neptune at 47.8 AU, was likely to endure for two separate reasons: 1) bodies captured into high-order mmrs during Neptune's outward migration, but now lying there, were at least semi-permanent members or, more specifically, a realistic model showed that only 2 bodies from a total of the 43 captured had escaped from 6 mmrs [from 6/13, at 50.4 AU, 5/11, 4/9, 3/7, 2/5, out to 1/3 at 62.6 AU] over integration times of 4.6 byr and 2) the many other outer belt bodies that were not in resonance, originated as escapers from the less retentive inner belt [a = or < a(1/2)] mmrs. This class, most often with large eccentricities, e, and inclinations, i, remained temporarily in this belt, likely to be replaced later or into the future by additional escapes from the same sources.

We now want to add that, while point 1) remains intact, 2) could use an extension because in Paper 1 we focused chiefly on escapes from the 1/2 resonance and not so much on other mmrs, in particular none of the other n/n+1 ones. These have an observational interest as some of them, 2/3, 3/4 and 4/5 have librating members, while as Fig. 1 indicates, still higher n/n+1 ratios do not. There is therefore some value in calculating their ability to capture bodies and to inquire into the lifetimes of those captured. We already know that questions of stability generally deal with an on-going process: many captured bodies later escape, reaching the outer belt and elsewhere. Here we approach stability by evaluating the degree of chaos in many mmrs, but we have not carried through very long term integrations.

By way of a summary, we can recap a few results from Paper 1 where sets of 500 bodies were placed for possible capture chiefly into mmrs between 1/2 and 1/3 as Neptune migrated by 7 AU over 10 myr. Of 69 with e(o) < 0.15 captured into 1/2, 31 escaped over 4.6 byr, with 10 of these doing so 'moderately recently', after 3 byr and 3 of those after 4.2 byr. Fully 2/3rds of the 38 remaining move in distinctly chaotic orbits, arguing that escapes will continue. Adding to the escapers from 1/2 were 9 others from a total of the 40 captured into higher inner belt mmrs, 3 each from 3/5 and 4/7 over times from 3.1 to 4.6 byr. Nearly all these escaping bodies occasionally moved into the region beyond a(1/2), for a very wide range of durations but averaging near 300 myr. It is on this basis that we concluded that the outer belt has been, and will in the future continue to be, well-stocked and that such bodies can eventually also make excursions into other parts of the solar system.

Our plan in this paper is to expand some of these efforts to include the other n/n+1

mmrs and especially 2/3 where the overall majority of librating bodies, now numbering about 200 in that mmr, are currently known. One goal is to account for the distribution in the inner Kuiper belt, a < a(1/2) and to check for other bodies likely to escape. To apply this step means folding in two other topics: 1) the likely distribution of KBO's before Neptune's migration and 2) estimates of capture probabilities.

Figures 1 and 2 begin this study by plotting the e's as a function of semimajor axis, a, from a = 32 to 66 AU of some of the 550 Kuiper belt bodies, observed for 4 or more oppositions, known in that region as of early 2013. Figure 1 shows the population in and around 6 first-order mmrs, 6/7 to 1/2, indicating an absence of members in the 5/6 mmr and no more than 2 [with e's > 0.3] in 6/7, and the observed eccentricity ranges of bodies in the four others. [Here we define a mmr by the mean motion ratio, KBO/Neptune.] For a comparison, Figs. 3 to 8 measure the degree of chaos in each of the six. These figures also indicate the e's where secondary resonances lie. The latter occur when the libration frequency is commensurate with a term in the apsidal frequency and an equivalence of the two, which corresponds to an overlap of two resonances, develops chaotic behavior. Circled crosses denote orbits that have escaped from a mmr, but we have not carried out a systematic survey looking for escapes from all of them, nor integrated most orbits beyond 500 myr. Integrations over the solar system's age for all orbits with Lyapunov times given by log T(L) < 3 or even 3.5 to establish a better link between chaos and escape must wait for another time. The Lyapunov time used here has been defined in earlier papers and is measured in orbital periods of Neptune. For the present we can assume with some assurance that orbits with log T(L) < 3 are unlikely to persist today.

A quick and slightly extrapolated summary based on Figs. 3 to 8 concludes that 1/2 can contain quite regular orbits in the range ~ 0.1 < e < 0.35 and that 2/3 shows regularity at e < 0.05 and between 0.1 < e < 0.33, but this resonance has definite chaotic regions 0.07 < e < 0.1 and beyond e ~ 0.33. At the higher e's stable librations are confined to amplitudes of only a few degrees. Two other mmrs, 3/4 and 4/5, show stable orbits up to e ~ 0.17 in the latter and to nearly 0.24 in the former, though both develop narrow chaotic zones at an individual secondary resonances. The ability of a mmr to prevent close approaches to the primary begins to fail for larger values of e. This is especially true at 5/6 where only a narrow e range up to ~ 0.07 corresponds to long lasting orbits. It was therefore a surprise to find a more extended stable area at 6/7, cf. Fig. 8, up to e ~ 0.14 but the 2 bodies near 6/7 in Fig. 1 with large e's must be only temporarily in resonance. At still higher ratios, 7/8 and 8/9 show no sign of regularity beyond e = 0.04.

This set of figures calls for two other remarks: 1) the effect of secondary resonances becomes most pronounced when the apsidal motion, dw/dt, is anticlockwise and 2) P(w) replaces P*(w) on Figs. 6 and 7 because here the secondary resonance is a commensurability between the libration period P*(l) and P(w) itself rather than a shorter period component, P*(w). Figure 2 also identifies unstable regions at 4 higher order mmrs in the outer belt, a > 47.8 AU, by providing values of e at which orbits show 2 levels of chaos.  Solid horizontal lines lie where log T(L) falls to 3 and the dashed one where it reaches 3.5.  The length of these lines, 1 AU, is a generous estimate of the total uncertainty of current semimajor axes, an estimate that also applies to bodies in Fig. 1.  Quite regular orbits can exist for e's less than the dashed marker.  In all cases in this study, we have probed to locate the least chaotic orbits.  That is, at any a and e, we could alter initial angular variables so as to find more chaotic, even colliding orbits, but we have plotted the most nearly regular ones.  For mmrs, this means searching for the most stable librations.  But there is no reason to suppose that all captured bodies will always find their way quickly into the most stable configuration.

**Some Conclusions from Figures 1 to 8**

The next paragraphs consider these figures after dividing them into two groups--the first one discusses the implications derivable from those with ratios equal to or higher than 4/5 plus a few remarks about 3/4, before then comparing 2/3 with 1/2. En route we shall introduce a new set of capture probabilities, P(c).  The absence of bodies at 5/6 and the higher mmrs might be [poorly] accounted for if the primordial belt began where Neptune's migration ceased to move the 5/6 mmr outward--therefore near 34 AU.  But rather than arbitrarily placing a supposed boundary that left very few objects for later capture, a sounder way is to start by considering capture probabilities at these higher ratios.  This plan is fueled by the knowledge that, while Jupiter readily captures at 2/1 and 3/2 in the asteroid belt [and equally well expels one-time captures from the former thanks to chaos owing to resonance overlap], the probability of capture at 4/3 is very low. We found in Paper 1 for Neptune and the Kuiper belt, that P(c) at 1/2 lies near 25%, for a migration time of 10 myr with the range depending somewhat upon the e(o)'s of bodies encountered. [Later when we compare 2/3 and 1/2, we consider also the finding that P(c)s may depend on migration time as well.]  For this paper, we again use an e-folding migration time of 10 myr and have obtained 2 sets of P(c)s for the following first-order mmrs, based on an integration of 400 bodies initially in the ranges 27.5 < a(o) < 39.5, 0.02 < e(o) < 0.08 and i(o)'s < 8 deg.

**Table I. Capture probabilities for a migration time, T(M) = 10 million yr.**

| Resonance | P(c) in %, plus number of bodies in ( ). | |
|---|---|---|
| | Initial Capture P(c,s) | Remaining after 200 myr P(c,l) |
| 6/7 | 7      (6) | <1    (1) |
| 5/6 | 3      (3) |        (0) |
| 4/5 | 9      (12) | 1     (2) |
| 3/4 | 9.7    (18) | 6     (12) |
| 2/3 | 18.6   (46) | 14.6  (36) |

Only for those captured into 2/3 have we extended the integrations to 1 and 2 byr, finding P(c) = 13% (33) for the longer time. In Table I the first set of P(c)s, labeled P(c,s), refers total number of captures, including the temporary ones that remained in a mmr for only up to 2 T(M) or 20 myr. Those with these short durations are usually the result of their e's being pumped upward to unstable levels. The second set, P(c,l), applies to those that remain captured at least to 200 myr, where these integrations stopped. Most of those that lie in 2/3 [31 of 36] move in regular or very mildly chaotic orbits, all with Lyapunov times measured by log T(L) > 3.7 and 14 of these have log T(L) > 4.5. Values as high as 4 are much rarer at 3/4 [3 of 12] and they are absent at 4/5. Figures 3-8 anticipate this behavior. For a partial comparison with the 1/2 mmr, we quote from Paper 1 for the limited sample with a slightly different range of e(o)'s, 0.04 < e(o) < 0.10, P(c)s of 31% and 21%, where the first applies to members initially captured and remaining for at least 200 myr and the second, 35 of the original 51, still present after 4.6 byr. We'll return to compare various populations a little later.

Bodies not found today in 5/6 and above, plus the reduced numbers observed in 4/5 and 3/4 relative to 2/3, are most likely the result of several factors: first the draining of the number of available candidates arising from the efficiency of prior capture opportunities into 1/2 and 2/3. [This is an effect taken into consideration when obtaining statistics.] But more to the point, Table I confirms that P(c)s obtained from a smooth initial distribution become smaller as n/n+1 -> 1 in some considerable part because even their short migration lifts eccentricities to unstable heights--that need rise above initial values only to 0.07 at 5/6 or > 0.14 at 6/7. These low limiting values of e are the real key denying long term permanence. Only by putting a clump of bodies positioned a bit smaller than their present locations-- i.e., just before the migration stopped--could a large current population be achieved. Given these factors, it seems a fair conclusion that they operate collectively to define the resulting inner boundary of the Kuiper belt once migration had ceased, whether it formally lies now at 5/6 [34.0 AU] or a bit smaller, but still 3 to 4 AU beyond Neptune's current orbit.

The 4/5 mmr with a small but measurable P(c) presents no surprises: it can accommodate e's up to 0.17 with some permanence, while all of its 7 known KBOs have equal or smaller values. Two other bodies with much larger e's, but still with a's near a(4/5), may correspond to ones acquiring such high e's from a long migration that they moved into a chaotic zone and eventual escape. It's a fairly frequent occurrence that bodies no longer librating in a mmr, having been formally ejected, remain for a time in its vicinity with only a small semimajor axis change.

The 3/4 mmr is populated up to e ~ 0.25 and a comparison with Fig. 5 leads to the likelihood that those with e > 0.25 have already escaped. The observed e distribution is readily reproduced to yield a continuous run of e's to 0.25 if either the initial distribution harbored e(o)'s up to 0.15 whence a short migration would increase e by > 0.10, or equally, capture occurred at a smaller e(o) earlier in the outward migration, since a(o) for 3/4 = 28 AU when Neptune moves by 7 AU.

The 2/3 mmr is an exciting and abundant place with a total of some 200 bodies, including also Pluto, lying within or very near this mmr.  A comparison of Fig. 4 with Fig. 1, where the former shows a chaotic region, owing to secondary resonance near e = 0.08, and an instance of escape from there at 330 myr, suggests a dynamical reason why the population in Fig. 1 at e < 0.10 is somewhat reduced.  The same comparison argues that escapes have diminished the number of bodies with e > or ~ 0.33 and also adds to the real possibility of future departures.

To turn now to other characteristics of 2/3 via a comparison with 1/2: Figs. 3 and 4 show some similarities but Paper 1 showed that at 1/2 the chaos is generated by the travelling of secondary resonances arising from variability in the libration frequency rather than changes in the apsidal terms and that this condition extends the chaotic region throughout the region, e < 0.1.  Severe chaos leads to escape in both for e's much above 0.3 at 2/3 and above 0.35 for 1/2.  At the same time there are differences between the stability diagram in Fig. 3 and the observations in Fig. 1.  First, an explanation for the paucity of bodies in Fig. 1 for e < 0.1 at 1/2 lies in a choice between dynamical instability acting on captured bodies or the absence of bodies available for capture.  Either option is likely enough--the second just requiring very few objects initially lying a few AUs less than a(1/2) now at 47.8.  But reasons why the population in 1/2 is much lower than 2/3 above e ~ 0.2 is not so easily settled.  To examine further, consider Table II that compares the currently known populations of these two mmrs and also of 2/5 [to be referred to later] over the constant range of de = 0.2 where reasonably regular orbits exist and also where the number of captures would be the same if we assume that the initial volume density was equal for all of them.

**Table II.  Observed populations at three mean motion resonances**

| mmr | a(o) and da for a 7 AU migration of Neptune | | e range | <Q> | N |
|---|---|---|---|---|---|
| 2/3 | 30.3 | 9.2 | 0.13 - 0.33 | 30.4 | 198 |
| 1/2 | 36.7 | 11.1 | 0.20 - 0.40 | 33.4 | 41 |
| 2/5 | 42.5 | 12.9 | 0.25 - 0.45 | 36.0 | 20 |

The N's are based on elements, available from the Minor Planet Center as of December, 2013, with a semimajor axis range corresponding to objects lying within

+/- 0.5 AU of the mmr.  This approximate figure covers both observational uncertainty and a measure of the normal amplitude of libration.

To prepare the N's for comparisons, we begin by making full use of capture probabilities, P(c), to predict an expected value for the 2/3 vs 1/2 ratio.  We have already mentioned a result from Paper 1 of a P(c) for 1/2 after 4.6 byr of 21% for the range $0.04 < e(o) < 0.10$ and T(M) = 10 myr.  In this paper we have obtained for 2/3 in the similar range, $0.02 < e(o) < 0.08$, values of 19%, and later 13%, first after 2 T(M) and then 2 byr.  The comparison is not exact, but we can reasonably assume that 1/2 is ~ half-again more likely to capture and retain that is 2/3, and therefore to determine their final population ratio, provided that both scanned through an initially equivalent particle field.  A prediction using just these P(c)s would then provide the result obtained by an observer situated in the outer solar aystem, equally distant from both of them.  For a terrestrial observer to predict what the locally measured ratio should be requires a correction because, relative to 2/3, a larger number remain 'undiscovered' in the more distant 1/2 resonance.  This correction depends on the <a>s and <e>s of the two and is given approximately by the 4th power of the mean perihelion distance where most discoveries are made, the Q's listed in Table II.  It turns out also to be ~ 1.5, arguing that the observed numbers for 1/2 relative to 2/3 should be reduced by ~ the same factor as had determined the former's higher P(c).

We are left with the clear conclusion that we should expect to see ~ the same number or maybe more in 1/2--a few more because its capturing region is more extensive [cf Table II]--when we make the equal encounter number assumption.  Table II's listing of the observed population ratio of 5 to 1 favoring 2/3 is so far different from the predicted one that assumptions need questioning.  A wise step would be to relax, or better to abandon, the provisional claim of a constant initial density of capturable bodies for both.  We can be sure, from Figs. 3 and 4, showing that levels of chaos in these two mmrs are so very similar in the $0.1 < e < 0.33$ range, that we cannot place any hope for an explanation relying on dynamical instability alone.

Two possible resolutions, acting together or separately, come to mind: 1) simply put, fewer bodies with e > 0.2 now in 1/2 could just reflect an initial distribution that tapers off before and then largely terminates near 46 AU or 2) the possibility investigated by Chiang and Jordan (2002) that the relative P(c)s for 1/2 and 2/3 depend critically on the migration time, T(M).  They determined that 1/2 is ~ twice as effective as 2/3 for T(M) = 10 myr, but the efficiency of 2/3 overwhelms 1/2 by at least a factor of 5 for T(M) = 1 myr.  These results apply to bodies captured and retained for 10 times T(M).  If we take the Chang and Jordan results at face value, their arguments imply that the corrected or true current 2/3 to 1/2 population ratio of ~ 3 to 1 would follow if T(M) is a few myr.  [N.B. The true ratio derived from the 'raw' observed one, does require the incompleteness correction be applied to 1/2.]

Table III begins a partial check on this claim by repeating the calculation of Table I

for T(M) = 1 myr:

**Table III. Capture probabilities for T(M) = 1 myr.**

P(c) in %, plus number of bodies in ( ).

| Resonance | Initial captures P(c,s) | Remaining after 200 myrs P(c,l) | Remarks |
|---|---|---|---|
| 6/7 | 4    (2) | 0 | |
| 5/6 | 11   (4) | 3    (1) | very chaotic |
| 4/5 | 7    (5) | 0 | |
| 3/4 | 15   (15) | 11   (11) | 6 regular |
| 2/3 | 11   (15) | 4    (5) | 3 regular |

[The details of Table III differ from those of Table I only as it draws upon just 200 bodies, but the P(c,l)s also extend for 200 myr.]

A comparison of Tables I and III leads first to a likely result that P(c) at 2/3 falls by a factor of 2 or 3 when T(M) drops to 1 myr, while in Paper 1 we found a decline in P(c,l) at 1/2 but by no more than 30% when T(M) was reduced from 10 to [only] 3 myr. Despite the approximate nature of this comparison, it still leads us to doubt whether lowering T(M) in the range from 10 to 1 myr is the sole contributor to the goal of greatly reducing the population ratio of 1/2 to 2/3. A caveat remains since integrations in this paper do not extend over the solar system's age. Another well populated mmr is 2/5 at 55.4 AU, currently with 20 likely members, about one-half as many as 1/2. At one time we planned to include possible inferences from it that might complement this discussion. We finally decided against doing so for the following reason: the 1st order 2/3 and 1/2 mmrs react to the eccentricity dependence of possible captures very similarly, both efficiently capturing bodies with e(o) < 0.1 and showing a distaste for ones with e(o)'s > 0.1. Higher order mmrs like the 3rd order 2/5 show a greater preference for e(o) > 0.1  This characteristic makes any conclusion sensitive to the initial e distribution, or leaves any comparison of 1/2 and 2/5 with too wide a range of possibilities to be really instructive, though the observed ratio of two to one, which requires no incompleteness correction, is a broadly consistent one.

We suggest then that both the details of the initial particle distribution and a migration time somewhat shorter than 10 myr could lead to a definite shortage of KBOs at 1/2. But we do claim that a falling initial population density after ~ 42 AU, dropping to very low levels after ~ 46 AU, is well established, a result that is also vigorously endorsed by the lack of bodies with e < 0.25, a limit that rises with semimajor axis up to ~ 0.35, in all mmrs situated beyond 50 AU. More quantitative progress needs a considerable number of detailed simulations that vary initial elements, the spatial distribution of bodies, migration times in the 1 to 10 mmr range and characteristics of the migration mechanism itself.

These lengthy remarks raise another point: the abundant nest of objects from 42.5 < a < 44.5 AU, conspicuous in Fig. 1, would have provided a wealth of candidates for 1/2 during its migration through that region, a passage that would have filled 1/2, 2/5 and others up to e ~ 0.2 and beyond. The current scarcity of KBOs in this part of 1/2 makes it likely enough that this populous grouping concentrated near <a> = 44 AU probably owes its origin to event(s) occurring either after or near the end of the migration--perhaps a family derived from a fragmenting collision, that could yield a large number of low e and i bodies. In any event, it need not necessarily be taken by itself as pertaining to features of the pre-migration solar system.

The tiny current population of non-resonant objects with a < a(2/3) raises another question. Table I of P(c)s shows that the preliminary percentage of captures into 2/3, P(c,s), is about 19% for T(M) = 10 myr, with other mmrs at even smaller values. The capture efficiency may depend on T(M), but it is still a fair question: what became of the remaining majority that were left behind, apparently never captured? Simulations provide a quantitative reply: during migration almost all such bodies were at least briefly captured or perturbed by higher order mmrs so that they would spend a short time affected by one mmr, then escaping with orbital parameters altered by enough to be subject to the effects of another, again resulting in a temporary capture. This cascading into and out of several mmrs occurs over times of a few thousand to hundreds of thousands of years. It might be called a scattering that slowly increases the initial e's and i's of a body to the point where it leaves the region 30 < a < 40 AU. It's this repeated behavior that also tends to develop inclinations, leading to the population frequently referred to as "the scattered disk". Our simulations indicate that <i>s of 20 to 25 deg. occur quite frequently and ones as high as 50 to 60 deg. occasionally, with such bodies regularly moving out, often temporarily, beyond a(1/2). This is an effective process but it does have a limited number of exceptions: about 8% [31 of 400] of those in the 10 T(M) integration with objects lying in the range 27.5 < a(o) < 39.5 AU were left undisturbed, with their e(o)s and i(o)s remaining essentially unchanged. Close to 65% of these had a(o)s = or > 36.5 AU, a fact that may be a consequence of the assumed e-folding exponential representation of migration--the 2/3 mmr travels ~ linearly from 30.3 to 36 AU. But this 65% is of some interest as Fig. 1 shows the presence of a number of low e bodies, probably not in any resonance, at least not in 2/3, from 37 to ~ 39 AU. However, for the vast majority of primordial bodies and especially those with a(o) < 37 AU, it is not an exaggeration to say that really the best way to preserve the identities of the initial population in this region was for its members to have been incorporated via capture into a very low order mmr, especially 2/3.

Included in Fig. 1 are three higher order mmrs, 5/8, 3/5 and 4/7 where, at least for the latter two, a fair number of bodies appears to be trapped. We have therefore considered their stability in Fig. 9. The case of 4/7 shows evidence of a secondary resonance near e = 0.1 but, apart from that, all three show no sign of marked chaos until e rises above ~ 0.25.  The two, 3/5 and 4/7, seem quite well populated, comparably so, and Paper 1 obtained their P(c)s as 15 and 7%.  The large number at small e in the higher order 4/7 mmr may well be another consequence of a later event mentioned earlier.  This argument for the seeming overpopulation of 4/7 compared to 5/8 is also reinforced by the fact that both these 3rd order mmrs have similar P(c)s.  In the integration of 400 bodies we did note 3 captures into regular orbits at 4/7 and a few more into regular ones into 5/7, 9/13 and even 11/16  In our earlier paper on the outer belt we noted two high order mmrs, 5/11 and 6/13, that captured and retained a few bodies over the age of the solar system.  With this in mind, apparent concentrations evident in Fig. 1, in the region a < a(1/2), at 6/11 [45.07 AU], 7/13 [45.46] and even 8/15 [45.75] seem quite believable.

KBOs located in the 4 mmrs beyond 1/2 shown in Fig. 2 provide added insights.  First we have found that, although all show severe chaos for e > 0.4, 3 of 4 do exhibit quite regular orbits for e's to ~ 0.40 and for the 4th, 3/7, as high as 0.35.  This fact and the complete absence of bodies with e's less than ~ 0.3 in 2/5 strongly implies that none were ever available for capture in the region outwards of ~ 46 AU.  We can elaborate on and reinforce this point by expanding some Paper 1 results which found that all 19 bodies captured into 2/5 remained librating there for 4.6 byr--no escapes even among those with small e's or also among the more chaotic examples.  Indicators of chaos for the 19 in that integration are:

| log T(L) | > 4.5 | 4.5 - 3.5 | < 3.5 |
|---|---|---|---|
| No. of bodies | 6 | 8 | 5 |

Their mean e's ranged between 0.14 and 0.40, averages that were well-defined for the more regular ones, but where individual values occasionally dipped to ~ 0 for about half the group.  Asking where a version of those hypothetical bodies of low e would now lie on a plot like Fig. 2 leads to the quite firm conclusion that the pre-migration distribution was one that could not supply any low e, e < 0.3, specimens and therefore that the very early outer boundary lay ~ 46 AU.

We can apply a similar argument at 1/2 from details in Paper 1, with more numbers but to a lesser effect.  There, of the 69 captures, 38 remained after 4.6 byr.  Eight moved in ~ regular orbits with log T(L) > 4.5 ; the remainder showed various amounts of chaos with the lowest log T(L) = 3.1.  The eight were characterized by 0.24 < e < 0.30 and the larger group by <e> between 0.09 and 0.30, all with e's sometimes dropping to ~ zero.  The observed population at 1/2 up to almost e = 0.2 is quite sparse, once again suggesting that few objects were ever present beyond ~ 46 AU.

One final example complements this picture.  At the 1/3 mmr Fig. 2 indicates the presence of a few likely members with e's up to 0.42, arguing for an increase, de, of at least 0.3 from pre-migration values of e(o) maybe as large as 0.15
[A reason for introducing an e(o) as large as 0.15 arises as Paper 1 found that mmrs of order > 1 surprisingly show equal or higher P(c)s in the e(o) range above 0.1 as for values below it.] Figure 10 shows that de = 0.3 implies that Neptune's migration must have driven 1/3 by ~ 16 AU. [This number may also be estimated from the approximate solution from the Lagrange planetary equations.] So large a distance means that these high e bodies were gathered up at a(o)s lying < 46.5 AU. Moving 1/3 by 16 AU requires that Neptune migrated by 7.7 AU. Using 2/5 as an example provides much the same result, though asking for the extreme value, de = 0.4, implies that Neptune moved by almost 10 AU, probably an upper limit.

## Concluding Remarks

The efforts described here began some time ago with a plan just to examine chaos in mean motion resonances and to look for the presence and possible influence of secondary resonances. Once some results were in hand, we decided to see how, in cooperation with other effects, they might bear on the evolution and distribution of bodies in the Kuiper belt even though other studies have already done much to clarify and interpret them. A few concluding remarks follow: first it is quite striking how well plots of the degree of chaos vs eccentricity shown in Figs. 3-9 match the observed e distribution at the mmrs, arguing that many-to-most escapes were a phenomenon of the past, or that once migration drives e's to a critical level, escapes will eventually occur. This situation implies that recent and future escapes will happen with decreasing frequency and/or that current ones result mainly from perturbations that transform a critical libration into an unstable one. Put differently: because the average time of outer belt residency is ~ 1/10 of the solar system's age [cf Paper 1] something like this type of replenishing is called for.

Our plan to map the secondary resonances within mmrs in the belt has proved rather unrewarding--they are present and identifiable but the chaos they generate has a minimal impact on the observed population. In a sense, it is unfortunate that where their influence in 1/2 at small e is most demonstrable, evidence we have gathered strongly suggests that the initial distribution provided too few candidates for capture. As we have already emphasized, the truncation of the initial distribution inferred from assorted mmrs and especially from 1/2, has made a major contribution to an explanation for the near complete absence of low e bodies in the outer belt resonances. For shorter migration times, the current population probably reflects a cooperation between the dependence of capture probability on T(M) and the nature of the initial distribution as the dominant factors.

A last comment concerns the extent of the Kuiper belt. Since the mmrs 5/6, 6/7,... can only scatter but not capture bodies with any degree of permanence, they locate, really force, the current inner boundary of the belt to lie near 34 AU. At the other end, the reduced membership in 1/2 and the eccentricity distribution in mmrs farther out, specifically the increasing e's below which very few objects have been discovered, help to place the initial outer boundary for capturable objects close to 46 AU.  Unlike the inner boundary, this one would seemingly pre-date any migration and so serve as a definition of the outer extent of asteroidal size bodies in the very early solar system.

**Figure Captions**

Fig. 1. Distribution of KBO's between 32 < a < 49 AU as of spring 2013, with principal mean motion resonances marked. We suspect that much of the concentration of bodies at/near 4/7 at 43.7 AU may be due to a post-migration collision.

Fig. 2. A continuation of Fig. 1 to 49 < a < 66 AU. Solid or dashed lines measure e's above which integrations and/or Lyapunov times indicate that escapes are either quite certain or may be possible.

Fig. 3. Chaos occurring at the 1/2 resonance from simulations of 400 bodies, e(o) < 0.08, i(o) < 8 deg. Circled crosses indicate cases that escaped in times < 500 myr. All crosses correspond to the least chaotic orbit, found by varying orbital parameters. P*'s denote either the libration period or the short period term in apsidal motion. Where their values are equal or commensurate, chaos clearly develops.

Fig. 4. Chaotic behavior at the 2/3 resonance from the same simulations, The 'cw' and 'acw' labels here and elsewhere denote clockwise of anticlockwise direction of apsidal motion. The 1:1 secondary resonance near e = 0.08 is particularly severe, with an escape noted at 330 myr.

Figs. 5-8. Chaos at 3/4 to 6/7. Taken as a group, Figs. 3-8 provide a survey of 6 1st order resonances, indicating where dynamically stable orbits occur and therefore where KBO's may--or may not--be expected to persist. P(w) replaces P*(w) when a commensurability involves the apsidal motion itself, not a shorter period term.

Fig. 9. Chaos at 3 well-populated higher order resonances in the inner belt.

Fig. 10. Eccentricity increase as a function of semimajor axis during a migration. Crosses are from the 2/5 resonance, darkened squares from 1/3.

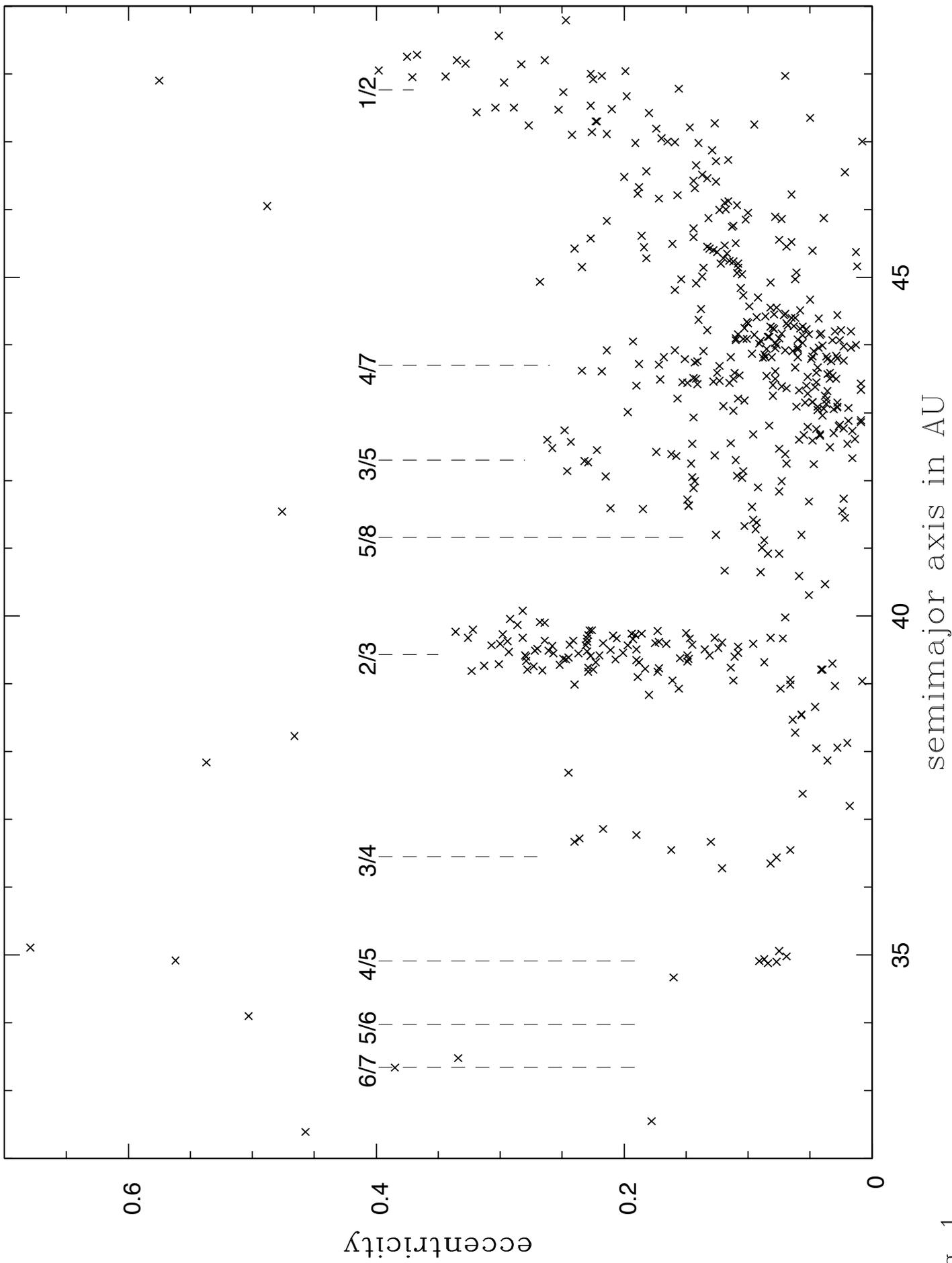

Fig. 1

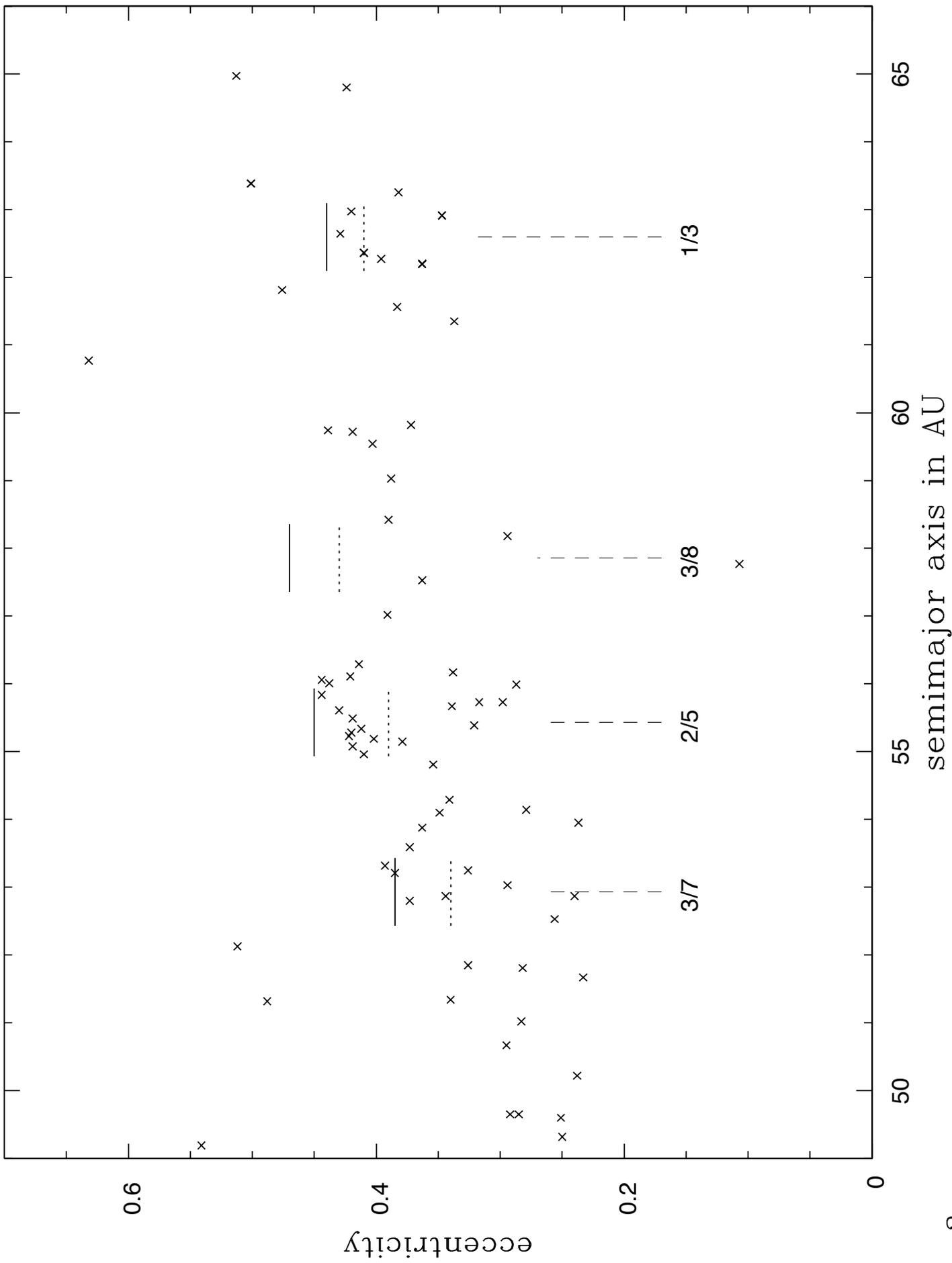

Fig. 2

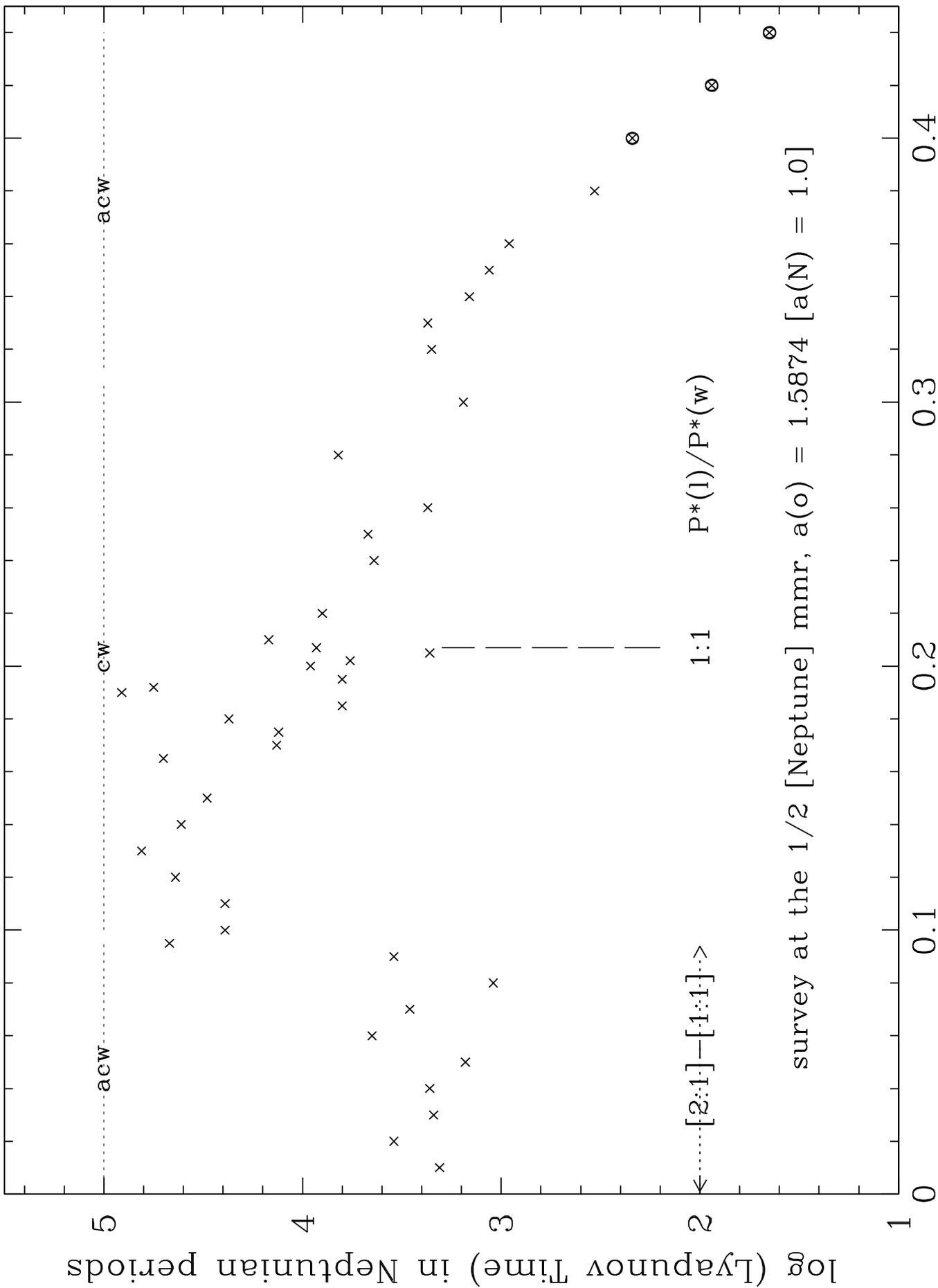

Fig. 3

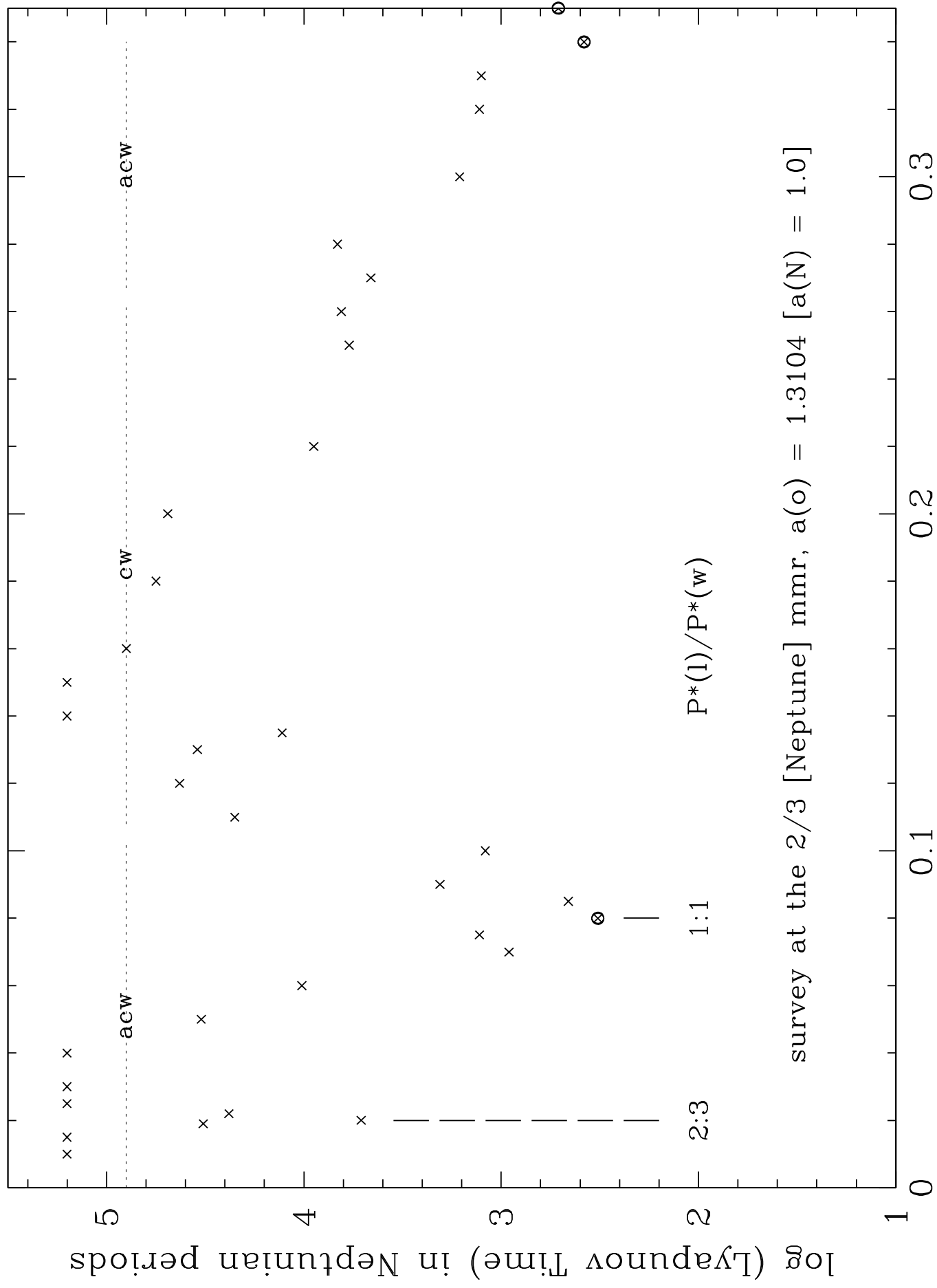
Fig. 4

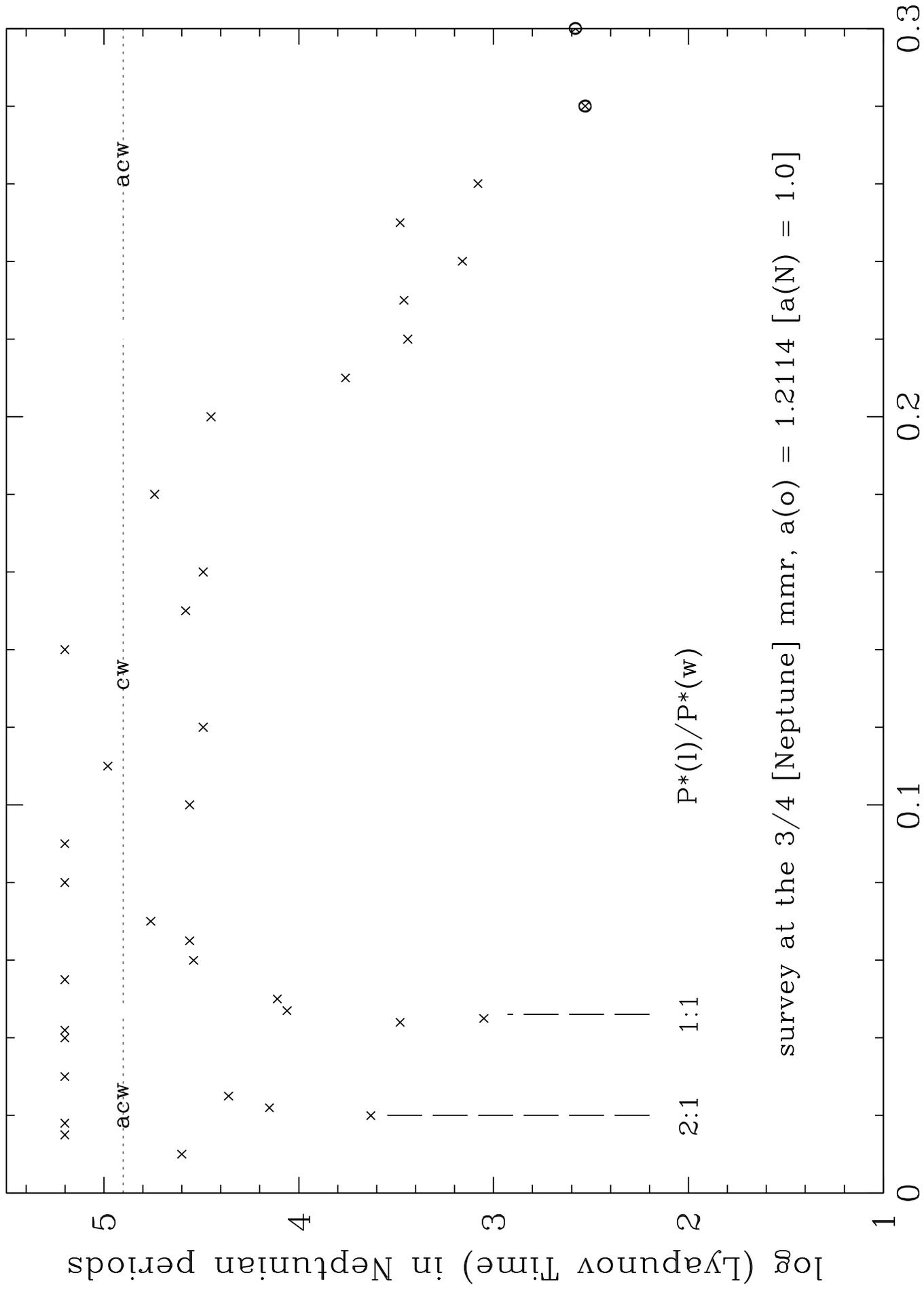

Fig. 5

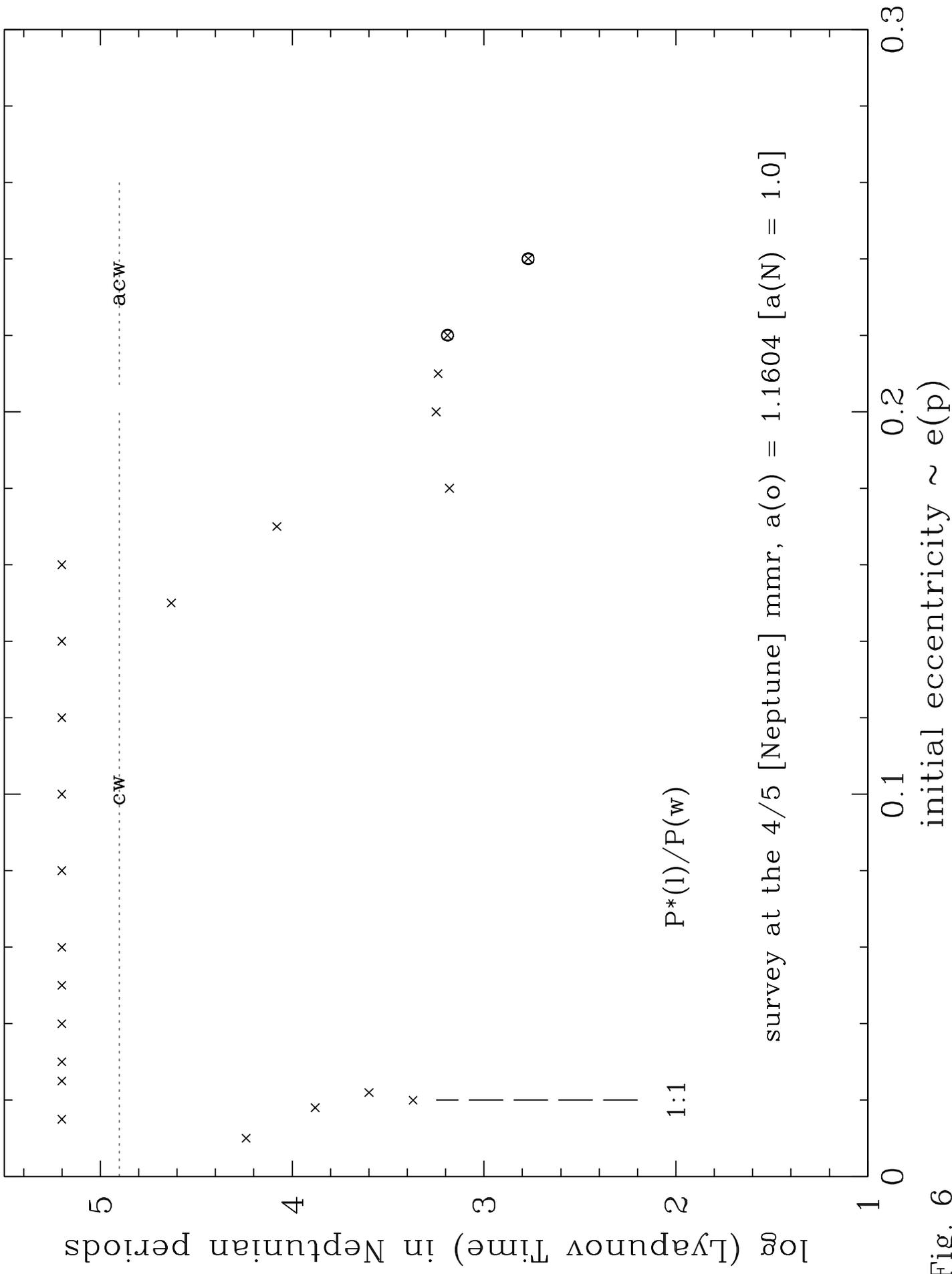

Fig. 6

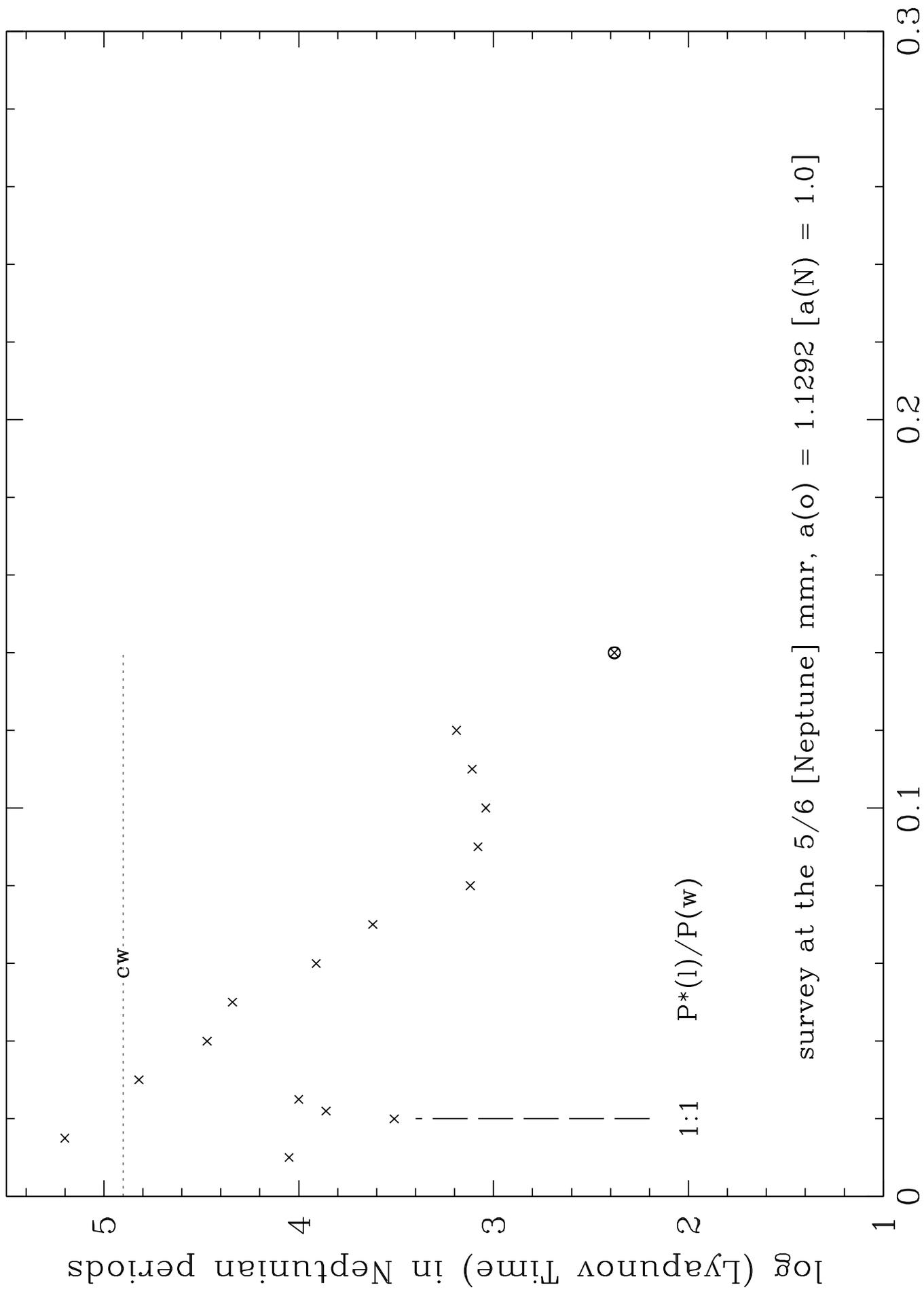

Fig. 7

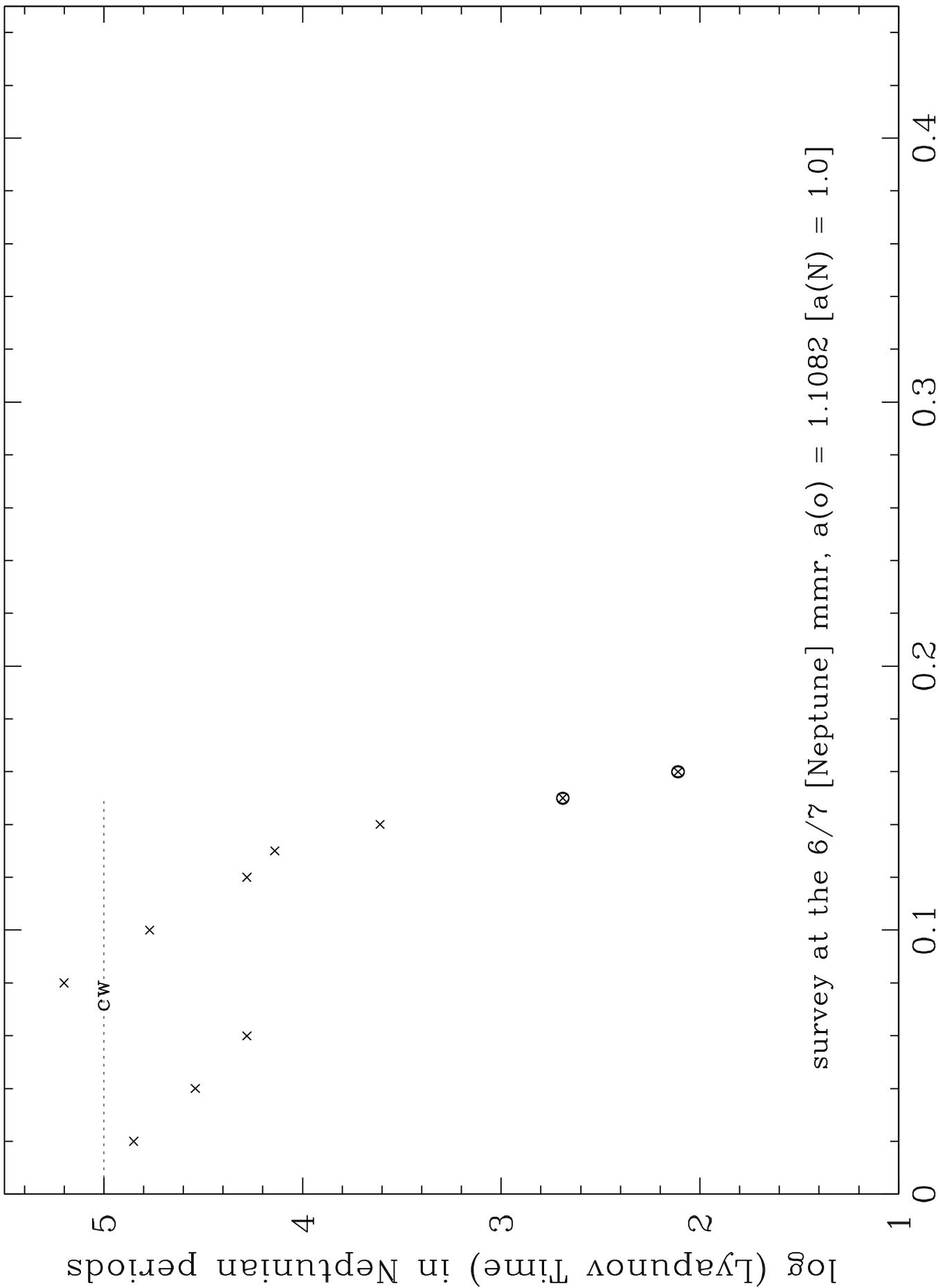

Fig. 8

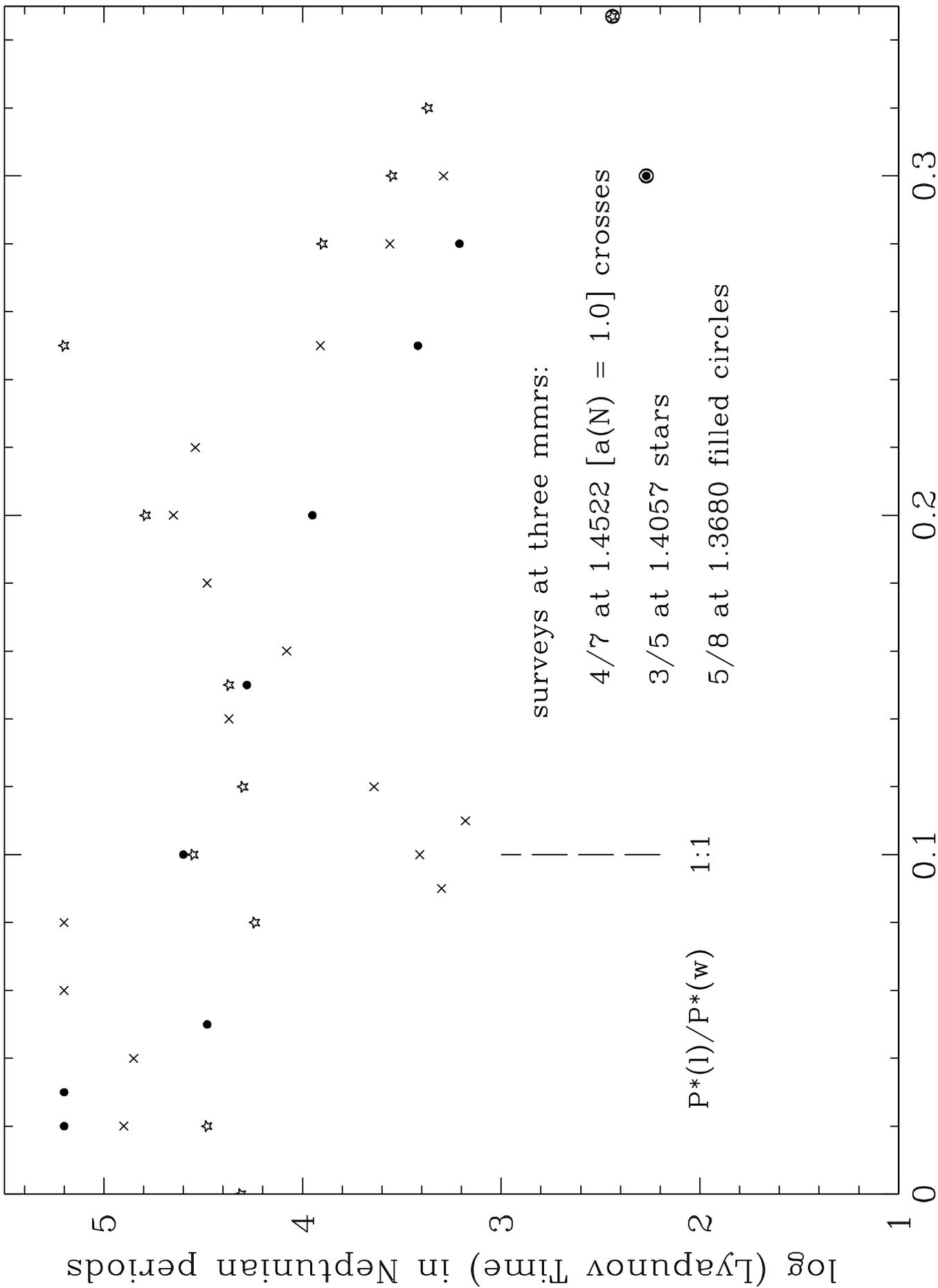

Fig. 9

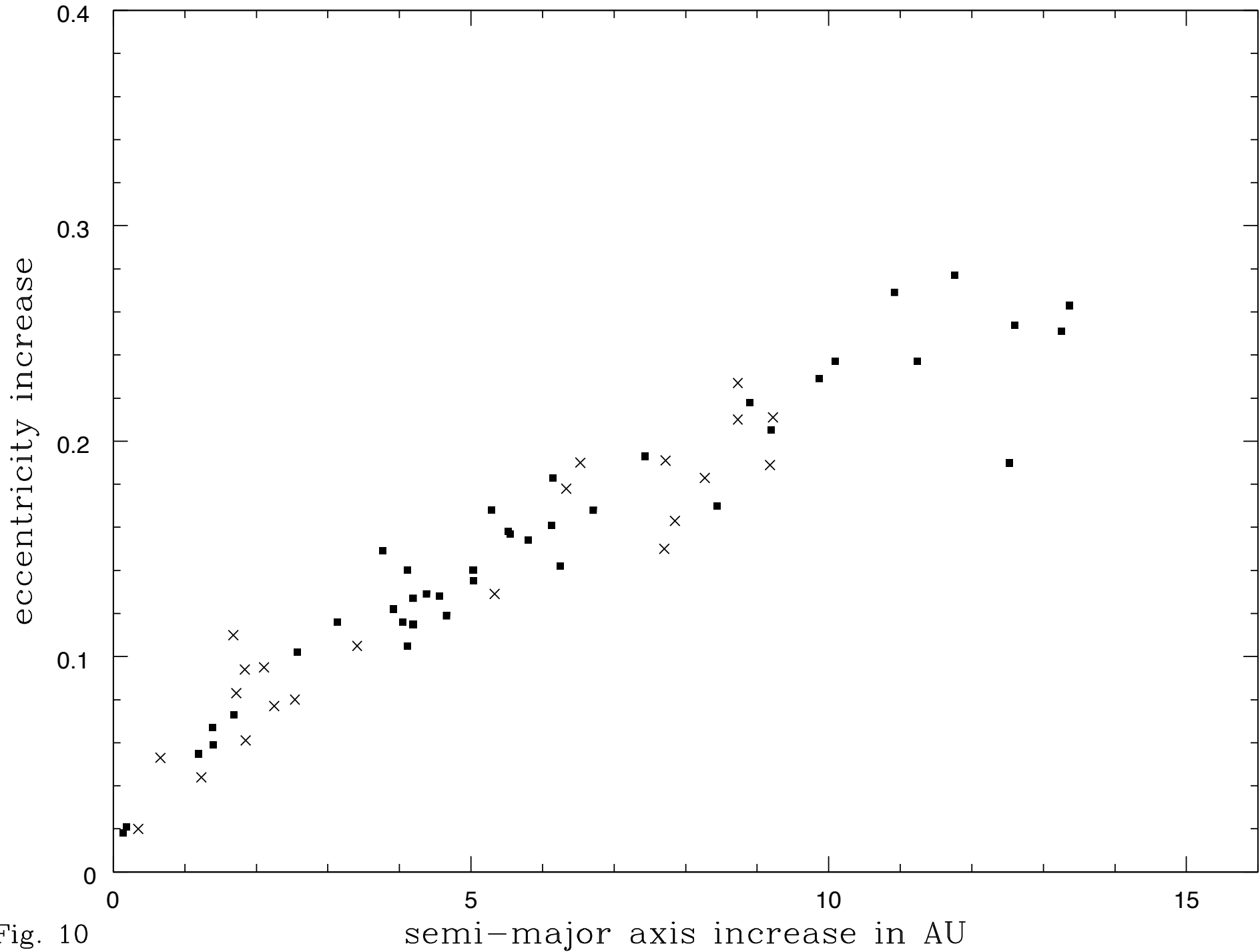

Fig. 10 semi-major axis increase in AU